\documentstyle[11pt,newpasp,twoside,epsf]{article}
\def\edcomment#1{\iffalse\marginpar{\raggedright\sl#1\/}\else\relax\fi}
\reversemarginpar

\newcommand{\Msun}{\mbox{$M_\odot$}}

\newcommand{\vi}{\mbox{$V\!-\!I$}}
\newcommand{\vk}{\mbox{$V\!-\!K$}}
\newcommand{\Teff}{\mbox{$T_{\rm eff}$}}
\newcommand{\feh}{\mbox{$[{\rm Fe/H}]$}}
\newcommand{\comment}[1]{}

\begin{document}
\title{Broad-band colour evolution of star clusters}
\author{L\'eo Girardi}
\affil{Dipartimento di Astronomia, Universit\`a di Padova\\
           Vicolo dell'Osservatorio 5, I-35122 Padova, Italy}

\begin{abstract}
We briefly review the main features in the broad-band colour
evolution of star clusters, over the complete age interval from 
$10^7$ to $10^{10}$~yr. The emphasis is in the problem of age-dating 
distant young clusters ($\la2$~Gyr) from their integrated colours. 
It is shown that \ub\ and \bv\ are less sensitive to metallicity than 
colours involving red pass-bands, like \vi, at least up to 
ages of some few Gyr. Since \ub\ and \bv\ are determined by 
well-understood and well-populated evolutionary stages, they are also 
less affected by theoretical uncertainties and by the ubiquitous 
effect of stochastic colour dispersion. The latter effects 
become important for the \vk\ colour. Thus, we argue that \ub\ and \bv\ 
are, presently, the more suitable broad-band colours for age-dating 
distant clusters. For other potentially useful colours like \vr\ and 
\vi, empirical tests of their evolution are still missing.
\end{abstract}


\section{Introduction}
\label{sec_intro}

The basic properties of distant star clusters, like 
ages, metallicities, or masses, can be derived only from integrated 
spectral properties. The obtention of spectra with reasonable S/N 
ratio is, in general, proibitive for galaxies located outside the 
Local Group (but for some few expections, e.g.\ NGC~7252; Schweizer 
\& Seitzer 1998). It follows that {\em broad-band magnitudes and 
colours are the primary source of information about masses and ages 
of distant star clusters}. Up to recently, this was true even 
for the relatively nearby Magellanic Cloud (MC) clusters.
We point out that from a total of about 2400 LMC clusters,
only about 300 have ages derived directly from the features in the
colour--magnitude diagram (CMD), whereas 600 have them estimated from 
the colours (Bica et al.\ 1996; Girardi et al.\ 1995). This latter 
situation is however rapidly changing, as massive and 
high-quality photometric data become available for the MCs.

In most papers in the field, the ages of distant star clusters are
estimated by a simple comparison of the observed colours with those 
predicted by a set of population synthesis models. As simple and
easy-to-apply this approach is, it is not demonstratedly the best 
one. In fact, for the nearby LMC clusters, empirical or 
semi-empirical methods have so far been preferred, and we are not 
aware of recent works in which the LMC clusters are age-dated from a 
direct comparison between model and observed colours. Some of the 
reasons for this will be mentioned in the following.

\section{Theoretical grounds}
\label{sec_theory}

\begin{figure}
\begin{minipage}{0.4\textwidth}
\plotone{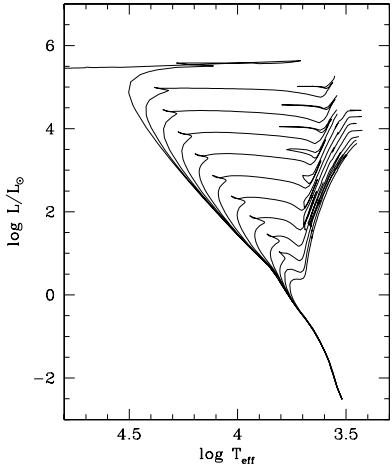}
\end{minipage}\hfill
\begin{minipage}{0.6\textwidth}
\caption{Evolution of theoretical isochrones for solar metallicities 
($Z=0.019$). They are derived from the Girardi et al.\ (2000; 
for $M\le7$~\Msun) and Bertelli et al.\ (1994; for $M>7$~\Msun) 
databases of stellar tracks. The TP-AGB phase is from Marigo et al.\ 
(1999). From above to below, the $\log t$ values go from 6.66 to 
10.33 (with $t$ in yr), at intervals of $\Delta\log t=0.33$.}
\protect\label{fig_isoc}
\end{minipage}
\end{figure}

The HR diagram of Fig.~\ref{fig_isoc} shows a sequence of isochrones 
at equally-spaced intervals of the logarithm of the age, $\log t$. 
From this figure it is evident that both $\log L$ and $\log\Teff$ of 
the turn-off stars fade at an almost constant rate with $\log t$. 
This is a simple consequence of the mass--luminosity relation of main 
sequence (MS) stars being a power-law. Core-helium burning (CHeB) 
stars follow the same linear trend in $\log L$, but only up to ages 
$\sim1$~Gyr. These trends determine, to a large extent, the main 
features of the evolution of integrated magnitudes: if the initial 
mass function is also a power law (e.g.\ the Salpeter one), one
expects an almost constant rate of fading of magnitudes with $\log t$.

\begin{figure}
\plottwo{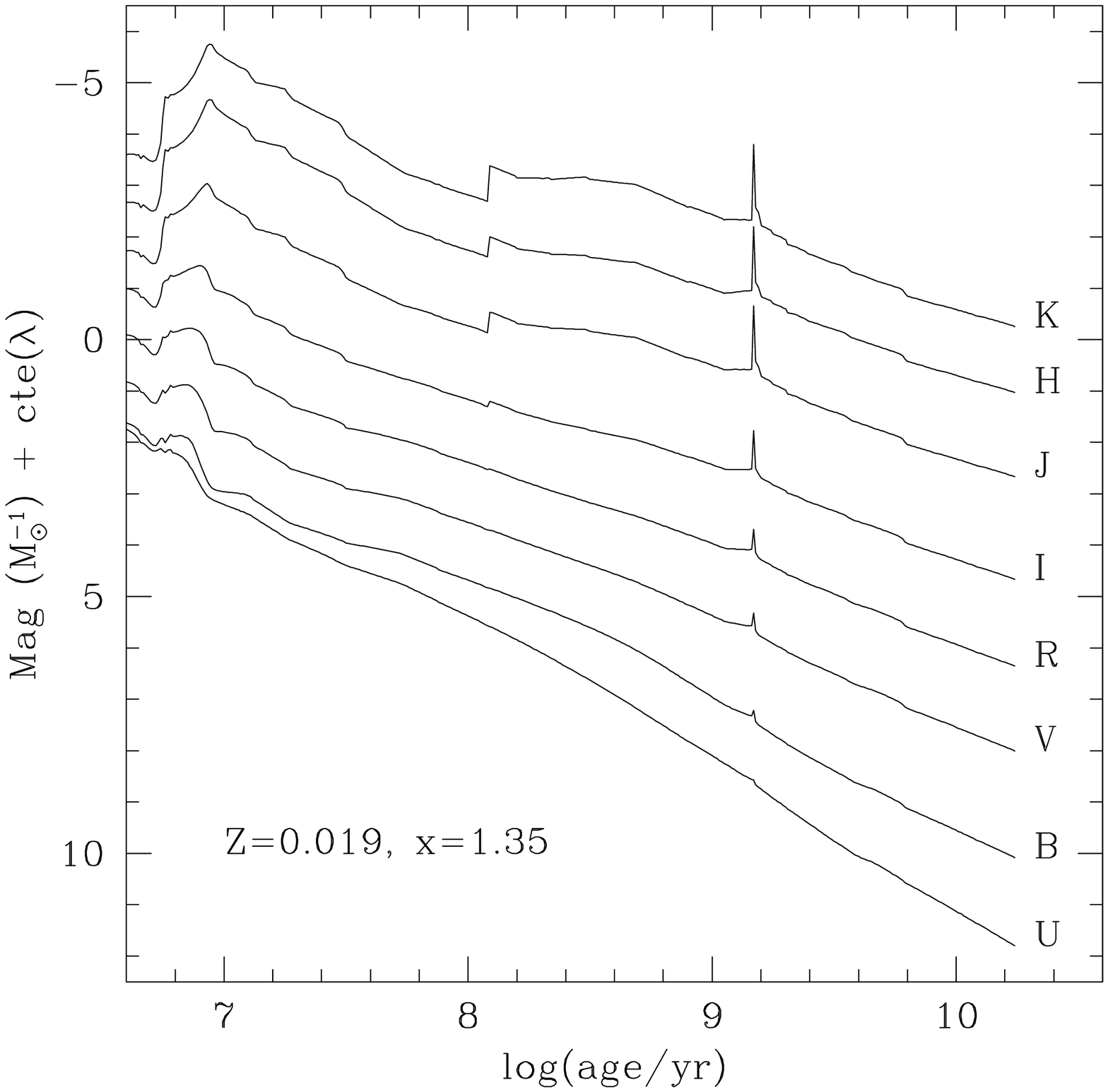}{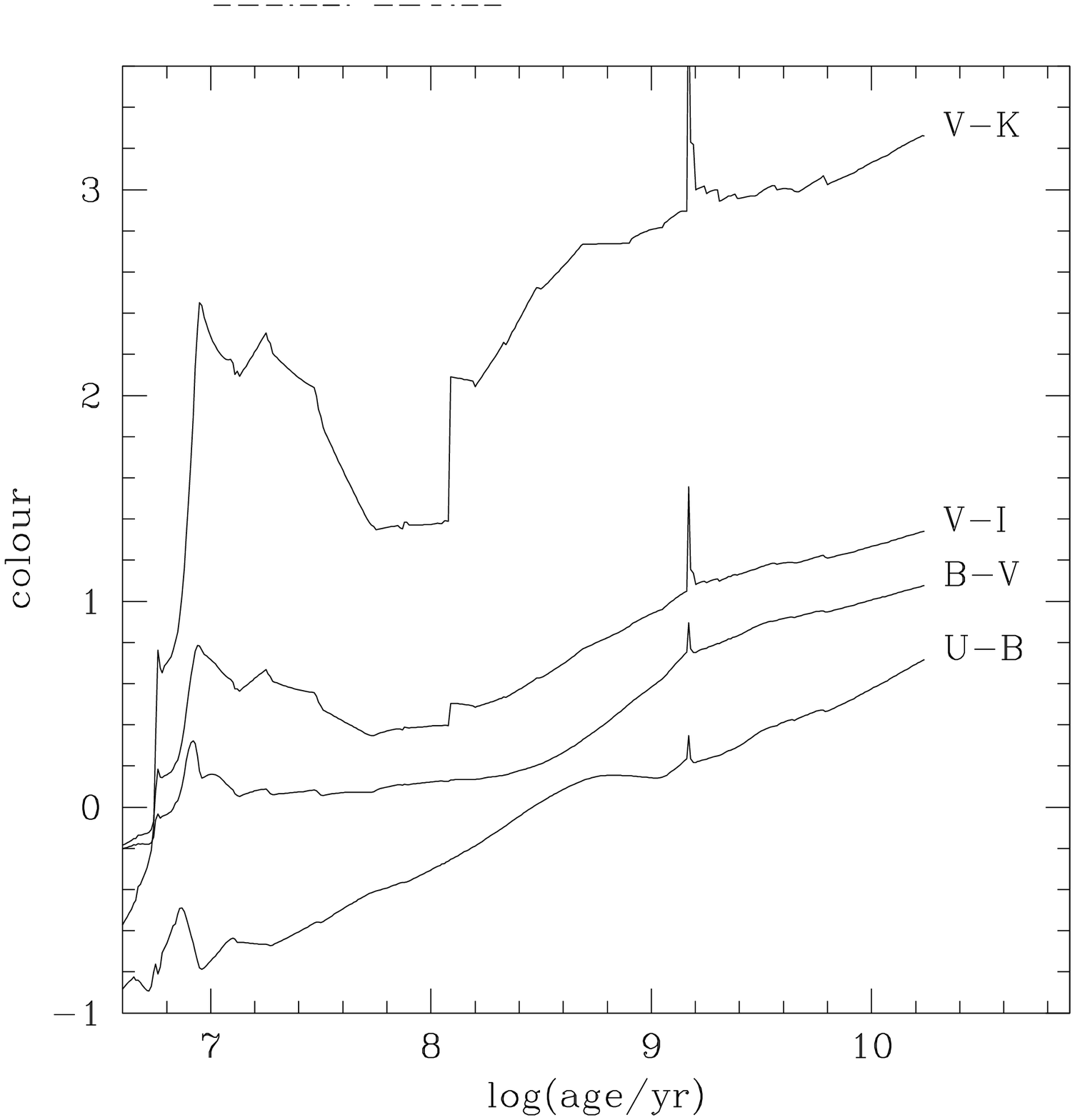}
\caption{Evolution of integrated broad-band magnitudes (left panel), 
and colours (right panel) with $\log t$, for $Z=0.019$, and a Salpeter 
IMF. The curves in the left panel have been shifted by integer 
quantities, in order to avoid their superposistion.
}
\label{fig_colour}
\end{figure}

And in fact, this is the case for the blue--visual pass-bands, like 
$UBV$, which depend mostly on the behaviour of stars in 
the stages of MS termination and CHeB. The roughly 
linear fading of $M_U$, $M_B$, $M_V$, and $M_R$ with $\log t$ (with 
slope $\sim2$~mag/dex) is shown in Fig.~\ref{fig_colour}. The same 
behaviour holds for the near-ultraviolet evolution (e.g.\ the 1550 
pass-band), but only up to an 
age of $\sim10^8$~yr; after that, turn-off stars are too cold to 
contribute to the UV spectra. 

The situation becomes different in the red and near-infrared. Evolved 
stars such as red supergiants (RSG), asymtotic giant branch (AGB)
and first-ascent red giants (RGB) determine the evolution at these
wavelengths. Since the number of these stars is a strong (and 
non-monotonic) function of $\log t$, the resulting magnitude and
colour evolution is also complicated. This can be seen in 
Fig.~\ref{fig_colour}: in $I$ and more clearly in $JHK$, the evolution 
is marked by the RSGs at $\ga10^7$~yr, and by the presence of AGB 
stars at $>10^8$~yr. Between $\log t\simeq8$ and $9.3$, the details 
of the colour evolution are essentially determined by AGB stars.
At $10^{9.2}$~yr, a transient red phase appears 
due to the increased number of AGB stars which temporarily follows 
the onset of the RGB (Girardi \& Bertelli 1998).

\section{Trends with metallicity}
\label{sec_metal}

\begin{figure}
\plottwo{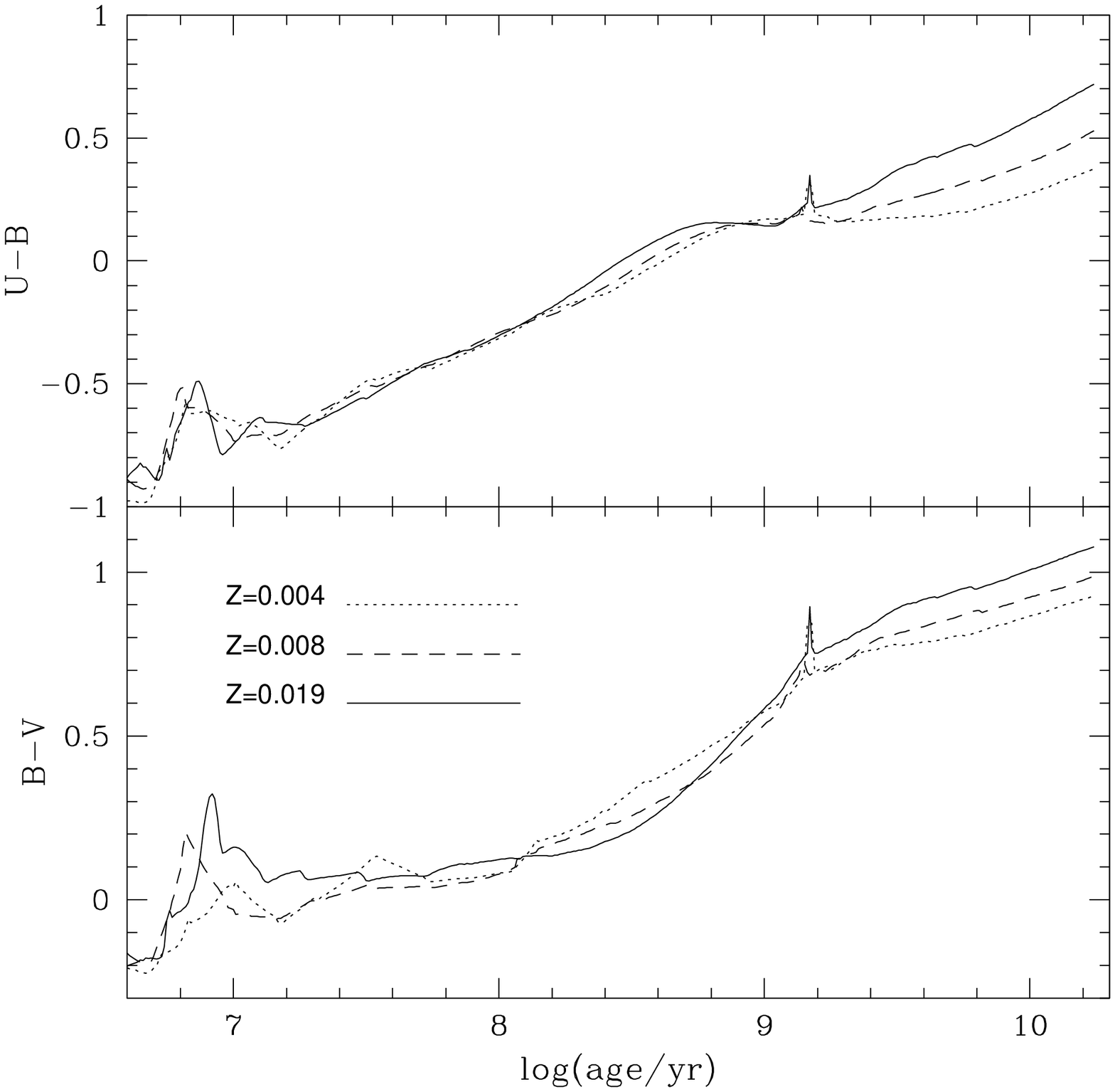}{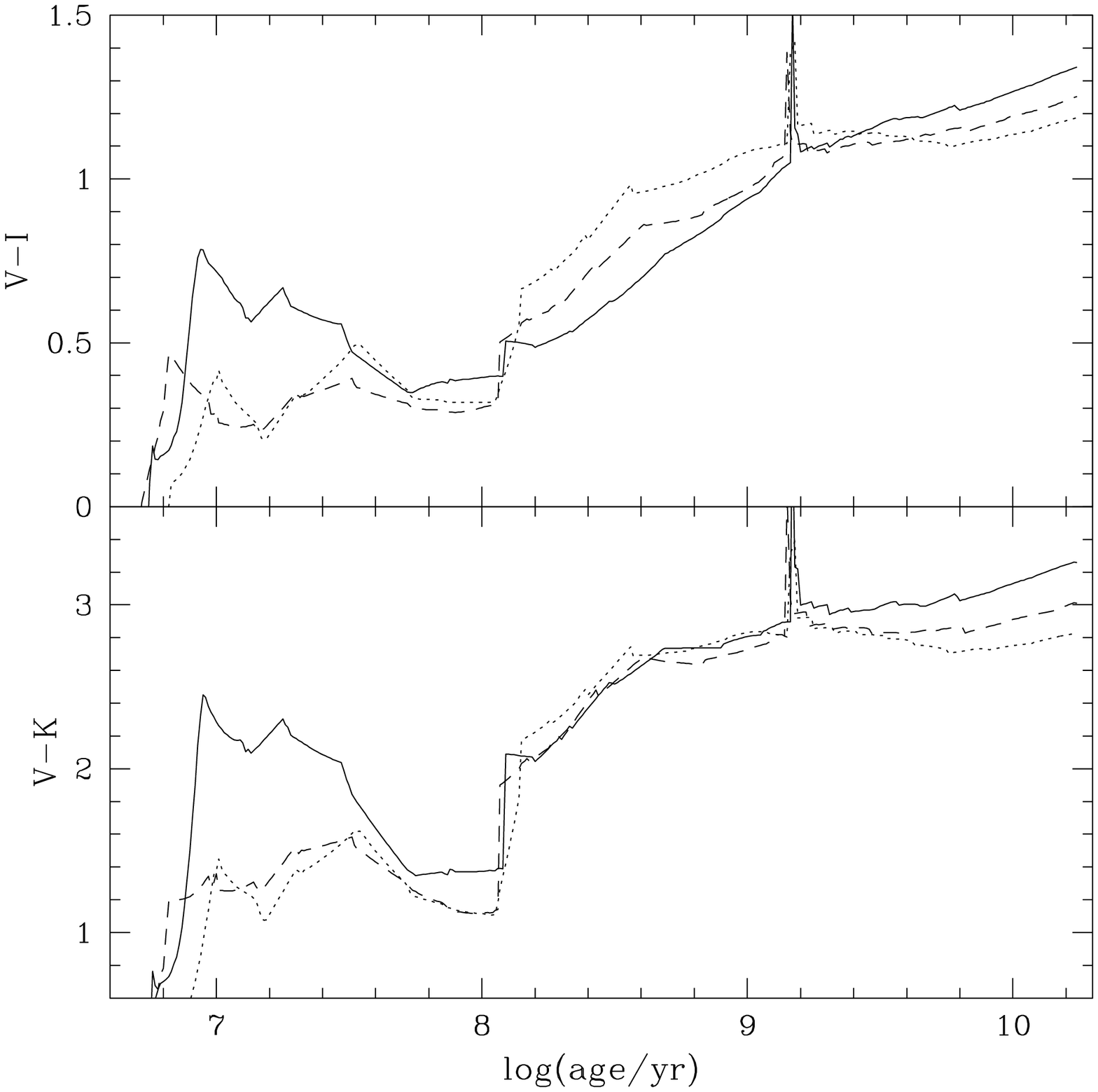}
\caption{Evolution of \ub, \bv\ (left panel), \vi, and \vk\ 
(right one), with $\log t$, for metallicities $Z=0.004$, 
0.008, and 0.019.}
\label{fig_metal}
\end{figure}

The above picture is valid for near-solar metallicities. Going to 
lower metallicities, some of the trends are just obvious: for the 
same $\log t$, colours get bluer. However, this effect is much 
larger at ages $\ge2$~Gyr, and for the red and near-infrared 
pass-bands. This is shown in Fig.~\ref{fig_metal}, with the colour 
evolution at different metallicities. For a large age interval, 
$7.3<\log t<9.2$, \ub\ and \bv\ colours differ by less than 
0.15~mag for metallicities in the interval $0.004<Z<0.019$. 
Moreover, the $\ub(t)$ and $\bv(t)$ relations are virtually 
monotonic. It follows that \ub\ and \bv\ are good age indicators, 
provided that $\bv\la0.6$ and $\ub\la0.2$.

The metallicity dependence of colours gets larger at ages 
$\log t>9.3$, when subgiant and RGB stars become important in 
comparison with MS and CHeB ones. This is the range of the 
age--metallicity degeneracy: ages can be estimated only if 
independent information on metallicity is provided, and vice-versa. 

At low metallicities, the RSG red phase is practically missing, and
AGB and RGB phases develop at higher temperatures. This result in 
small changes in the behaviour of \ub\ and \bv\ colours, which are,
fortunately, of low entity for most of the age interval, i.e.\
$7.3\la\log t\la9.2$. In the $IJHK$ pass-bands the changes 
become remarkable, since these stars are the main contributors to the 
red and near-infrared flux. It comes out that colours like \vi\ and 
\vk\ are very dependent on the way AGB stars are considered in the 
models (see Girardi \& Bertelli 1998). Moreover, the colour-age 
relation is not monotonic for \vi\ and \vk. Thus, when taken 
separately from other colours, they are not good age indicators.

A remark is worth here. It is often assumed that all 
models of population synthesis predict equally well the evolution of 
broad-band colours, since they are, in general, based on the same sets 
of evolutionary tracks and synthetic spectra. This is not exactly true. 
One important source of difference between models is on the way different 
groups consider the termally pulsing AGB phase. In some cases, TP-AGB 
stars are even neglected; this is a bad assumption since up to 50~\% 
of the bolometric luminosities, at certain ages, come from these 
stars (cf.\ Frogel et al.\ 1990).
In other cases, TP-AGB stars are included according to some empirical 
prescription, as e.g.\ that derived from the Frogel et al.\ (1990) 
observations of AGB stars in MC clusters. Although apparently better,
this latter approach does not include any metallicity dependence in
the TP-AGB evolution, whereas every single aspect of this evolution
(mass loss, dredge-up, hot-bottom burning, lifetimes) is expected to
be strongly dependent on metallicity\footnote{These dependences, even 
when not explicitly included in the equations of TP-AGB evolution, 
come naturally from the changes in the typical temperatures of AGB 
envelopes with metallicity.}. The best approach, probably, is 
to use TP-AGB models whose parameters have been calibrated in order 
to reproduce the properties of AGB stars in the MCs, like those by
Van der Hoek \& Groenewegen (1997) and Marigo et al.\ (1999). The
latter ones are adopted here. 

For near-solar metallicities, TP-AGB stars affect mainly
the red and near-infrared passbands (e.g.\ $IJHK$) and
colours like \vi\ and \vk. At low metallicities, TP-AGB
stars slightly affect also the visual colours, like \bv\ and \vr\ 
(see Girardi \& Bertelli 1998). These are the
colours and metallicity ranges in which different population
synthesis models may differ the most. We point out that
the metallicity dependence of \vk\ is almost absent in the 
$\log t>8$ models of Fig.~\ref{fig_metal}, just {\em because} 
the underlying TP-AGB models include a considerable dependence 
on this parameter, which compensates for the changes in the mean 
AGB temperature. Different trends would be found in other models. 

\section{The problem of age determination}
\label{sec_age}

A variety of effects, other than age and metallicity, determine the
integrated colours of star clusters. A very subtle and ubiquitous
one is that of ``stochastic fluctuations'' in the integrated colours 
due to the limited number of stars. The effects are sizeable already in 
\bv\ colours, even for the populous LMC clusters which have $M_V\la-6$. 
For $M_V=-6$, the standard deviation of \bv\ varies from 
$\sigma(\bv)\simeq0.25$ at $\log t=7.3$, to $\sigma(\bv)\simeq 0.03$ 
at $\log t=9.0$ (Girardi et al.\ 1995). It scales approximately as 
$\sigma(\bv)\propto0.7^{-M_V}$.\footnote{Notice that, in Eq.~3 of
Girardi et al.\ (1995), the exponent is presented with the wrong 
sign. The correct equation should read $\ldots\times0.7^{-(M_V+6)}$.}
Thus, only for intermediate-age clusters, and for young ones with 
$M_V\la-9$, is this intrinsic dispersion lower than that caused by 
cluster-to-cluster metallicity variations of 
$\sigma(\feh)\simeq0.3$~dex. 
 
In contrast, more dramatic stochastic fluctuations are found
in the red and near-infrared pass-bands, due to the sampling of a 
small number of luminous red stars per cluster (e.g.\ the AGB ones; 
see Santos \& Frogel 1998; and Lan\c con, this meeting).
Also in this respect, the blue-visual pass-bands do a better job 
than red and near-infrared passbands, since they sample the light 
coming from stars in well-populated evolutionary stages. 

To summarize, colours such as \bv\ and \ub\ present several potential
advantages for the age determination of young clusters, as the 
lower sensitivity to metallicity, stochastic flutuations, and their 
monothonic behaviour with age. In fact, the most classic way to 
determine the ages of MC clusters from colours, namely the Elson \& 
Fall (1985; EF) one, makes use of the \bv\ versus \ub\ two-colour 
plane. Clusters draw a fairly regular $S$-shaped curve in this plane, 
so that their position along this curve simply indicate their 
$\log t$ values. The EF method has been slightly modified by Girardi 
et al.\ (1995) in order to take into account the fact that the 3 most 
important factors of dispersion of colours at fixed age -- namely the 
metallicity, reddening, and stochastic dispersions, -- act in almost 
the same direction in the two-colour plane. However, it has been 
demonstrated that the age determinations with the EF method become 
unreliable for clusters older than say $4$~Gyr, due to the 
age--metallicity degeneracy above mentioned. 

The EF method has the advantage of being empirical, since 
the age sequence of clusters in the two-colour plane is calibrated 
using LMC clusters with ages directly determined from the 
CMD features. In this way, uncertainties in the integrated colours
of models (always present and of order $0.05-0.1$~mag in $UBV$) do 
not affect age estimates. Determinations of $\log t$ obtained this 
way present accuracies of $0.15$~dex, at least for the most 
populous LMC clusters (Girardi et al.\ 1995). We remark 
that, using a simple comparison between the LMC cluster colours and
population synthesis models, the typical accuracies would be much 
worse, i.e.\ $\sigma(\log t)\sim0.3$~dex. The validity of the EF
method has also been confirmed for SMC clusters by Grebel et al.\ 
(1999). 

Considering these aspects, it is surprising that 
empirically-calibrated methods are rarely used (and exceptions
are e.g.\ Bresolin et al.\ 1996; and Larsen \& Richtler 1999) for 
the age-dating of distant clusters. Moreover, very often are the 
ages of distant clusters derived from \vi\ or \vk, which, from a 
theoretical point of view, are not the most apropriated colours. 
To be more specific, the age 
information these colours contain is poorer than that contained in a 
single measure of \bv\ and/or \ub, because of the effects of 
non-monoticity with $\log t$, metallicity dependence, and/or 
stochastic dispersion. For \vk, the importance of these effects can 
be confirmed by simply looking at the way \vk\ colours change among 
clusters of similar age in the LMC (Santos \& Frogel 1998; Girardi 
\& Bertelli 1998). More compelling reasons for avoiding the red
pass-bands in age determinations, can be derived from population 
synthesis models. One finds that the predicted \vi\ and \vk\ colour 
evolution can differ appreciably from author to author, as a result 
of the different ways in which the several models include the 
TP-AGB evolutionary phase. This uncertainty is not present in the 
$UBV$ pass-bands. 

However, it is also true that most modern instrumentation, and 
the strong absorption met in several interesting galaxies, 
favour the observations in pass-bands like 
$VRIK$, instead of the $UBV$ necessary for easy age determinations. 
Measuring clusters in red pass-bands is, in many cases, a necessity.

The $BVRI$ pass-bands are, probably, those in which the best data 
for distant clusters are available. To improve upon the age 
estimates obtained from these data, it would be adviseable to test, 
empirically, the age and metallicity dependences of the \vr\ and 
\vi\ colours. We notice that this can be done, to some extent, 
using the present-day photometric databases for MC clusters (e.g.\ 
OGLE, MACHO, and the MC Photometric Survey; see the contributions 
in Chu et al.\ 1999). Only after these behaviours are well 
documented, will we be able to check the reliability of such age 
determinations. Also, this kind of work may constitute an 
important test to models of spectral evolution of stellar 
populations, regarding, especially, the way they include (or do 
not) the latest evolutionary phases for different metallicities.

\acknowledgements
We acknowledge support by the European Community under 
TMR grant ERBFMRX-CT96-0086, and by the Italian MURST.
Many thanks are due to G.\ Bertelli, E.\ Bica, A.\ Bressan, 
C.\ Chiosi, P.\ Marigo and A.\ Weiss for the many useful 
discussions and remarks.

\end{document}